\documentclass[preprint,showpacs,aps]{revtex4}

\usepackage{graphicx}
\usepackage{dcolumn}
\usepackage{bm}

\begin{document}

\begin{titlepage}

\title{Theoretical Study of Corundum as an Ideal Gate Dielectric Material for Graphene Transistors}

\author{Bing Huang, Qiang Xu, and Su-Huai Wei}

\address{National Renewable Energy Laboratory, 1617 Cole Boulevard, Golden, CO 80401, USA}

\date{\today}

\begin{abstract}

Using physical insights and advanced first-principles calculations,
we suggest that corundum ($\alpha$-Al$_{2}$O$_{3}$) is an ideal gate
dielectric material for graphene transistors. Clean interface exists
between graphene and Al-terminated (or hydroxylated) Al$_{2}$O$_{3}$
and the valence band offsets for these systems are large enough to
create injection barrier. Remarkably, a band gap of $\sim$ 180 meV
can be induced in graphene layer adsorbed on Al-terminated surface
with an electron effective mass of $\sim$ 8 $\times$ 10$^{-3}$
m$_{e}$, which could realize large ON/OFF ratio and high carrier
mobility in graphene transistors without additional band gap
engineering and significant reduction of transport properties.
Moreover, the band gaps of graphene/Al$_{2}$O$_{3}$ system could be
tuned by an external electric field for practical applications.

\end{abstract}


\maketitle

 \draft

\vspace{2mm}

\end{titlepage}

Graphene has been studied intensively due to its unique electronic
and mechanical properties such as extremely high carrier mobility
(over 200000 cm$^{2}$/V$\cdot$s for suspended
samples)\cite{Geim-Castro Neto}. However, to utilize graphene, which
has a zero band gap when it is pure, for electronic devices such as
the field-effect transistors (FETs), it is essential to open up a
band gap in graphene to realize the ON/OFF switch function. One way
to open the band gap in graphene is utilizing the quantum
confinement effect, e.g., etching graphene into one-dimensional
nanoribbons\cite{Qimin-2007, M. Y. Han-2007, X. Li-2008, Wang-2008}.
In practice, very narrow graphene nanoribbons (GNRs) ($\sim$ 10 nm)
are necessary to achieve a band gap of $\sim$ 0.2 eV (ON/OFF ratio
$>$ 10$^{2}$)\cite{Qimin-2007, M. Y. Han-2007, X. Li-2008,
Wang-2008}, but a large scale production of such narrow GNRs is
still quite challenging. Moreover, the carrier mobility of GNRs
(about hundreds of cm$^{2}$/V$\cdot$s\cite{X. Li-2008, Wang-2008})
is several magnitudes lower than that of graphene sheet due to the
(intrinsic) band folding and phonon scattering\cite{A. Betti-2011}
as well as (extrinsic) difficulty in controlling the edge quality in
experiments\cite{X. Li-2008, Wang-2008}. Another way to open a band
gap in graphene is breaking the inversion symmetry of the A, B
sublattices, e.g., by placing graphene onto some special substrate.
In this case, the band structure near the Dirac point or the carrier
mobility of the graphene is better preserved if the interaction
between graphene and substrates is weak. Obviously, this approach
has technological advantages over the etching of graphene. However,
a simple guideline on how to search an ideal substrate that could
induce a sufficiently large band gap in graphene is still unclear,
especially for substrate which can be integrated directly into the
current FET technology.

Most of the graphene FETs to date employ silicon oxide (SiO$_{2}$)
as the bottom-gate dielectric and an alternative
high-dielectric-constant (high-\emph{k}) material as the top-gate
dielectric (e.g., the amorphous structures of
HfO$_{2}$\cite{Zou-2010} and Al$_{2}$O$_{3}$\cite{Liao-Kim}). The
integration of a high-\emph{k} top-gate can push the FET performance
to a much higher limit because it can better screen charged
impurities and enhance carrier mobility in graphene\cite{C.
Jang-2008, Liao-Kim, Zou-2010}. Analogy to silicon FET, an optimal
high-\emph{k} gate dielectric material should have high dielectric
constant, large injection barrier (i.e., large band offset ($> 1$
eV) with respect to graphene), high chemical stability, and no or
minimal interface states at the high-\emph{k}-oxide/graphene
interface\cite{Xiang-2009}. In addition, an ideal high-\emph{k}
dielectric is desired to have the ability to open an adequate gap in
graphene for FET operation.

Although several calculations have shown that a small gap could be
induced in graphene by SiO$_{2}$\cite{Y. Kang-2008, N. T.
Cuong-2011}, HfO$_{2}$\cite{K. Kamiya-2011}, or BN\cite{G.
Giovannetti-2007} substrates, a general rule to find a substrate
which can induce a larger band gap in graphene is still lacked.
Here, we propose that one should search for the substrate which has
atoms with large chemical potential difference at the surface and is
commensurate with the graphene lattice. These two conditions can
cause large potential difference $\Delta_{AB}$ at the A, B
sublattices of graphene, thus produce large band gap in graphene.
Under this physical insight, we suggest that the (reconstructed)
Al-terminated $\alpha$-Al$_{2}$O$_{3}$ (0001) surface is an ideal
high-k-oxide substrate for graphene FETs. This is because after the
reconstruction both Al and O are present at the surface and the
chemical potential difference between Al and O is much larger than
that of other popular substrates such as SiO$_{2}$, BN, and
HfO$_{2}$ studied before, and the lattice mismatch between
$\alpha$-Al$_{2}$O$_{3}$ (0001) and graphene is also relatively
small. Carried out by advanced first-principles calculations, we
demonstrate that a large band gap of $\sim$ 180 meV at the Dirac
point appears in graphene layer adsorbed on Al-terminated (0001)
surface with a quite small effective mass of $\sim$ 8 $\times$
10$^{-3}$ m$_{e}$ for Dirac fermions. As we expected, the band gap
of graphene on Al$_{2}$O$_{3}$ is significantly larger than on other
widely used substrates\cite{note1-LDA}. The size of the band gaps of
graphene/Al$_{2}$O$_{3}$ systems could be further tuned by an
external electric field for practical applications. Moreover,
interface states can be eliminated if graphene is grown on
Al-terminated (or hydroxylated) Al$_{2}$O$_{3}$ (0001) surface and
the valence band offsets for these hybrid systems are large enough
for injection barrier.

All the density-functional-theory (DFT) calculations are performed
by using the VASP code\cite{VASP}. Projector augmented wave (PAW)
potentials are used to describe the core electrons, and generalized
gradient approximation (GGA) with the PBE functional is selected in
our calculations. We find that the van de Waals (vdW) interaction
plays an indispensable role in accurately determining the adsorption
configuration and binding strength in this system. The effect of vdW
interactions is taken into account by using the empirical correction
scheme of Grimme (DFT + D/PBE)\cite{vdW}, which has been proved to
be successful in describing the geometries of graphene related
structures\cite{G. Mercurio-D. Stradi}. Our DFT+D/PBE calculations
show that the interlayer distance of graphite is 3.22 \AA~ and the
interlayer binding energy is -54 meV/C-atom, in excellent agreement
with quantum Monte Carlo calculation (-56 meV/C-atom)\cite{L.
Spanu-2009} and experimental value (-52$\pm$5 meV/C-atom)\cite{R.
Zacharia-2004}. It is well-known that GGA type calculations usually
underestimate the band gaps of semiconductors and the absolute band
edge energy from the GGA calculation is not always reliable. Since
hybrid functional calculations could give improved results for both
conventional semiconductors\cite{A. Alkauskas-2008} and graphene
nanostructures\cite{V. Barone-2006}, we adopt HSE hybrid functional
to calculate the electronic structures of these
Al$_{2}$O$_{3}$/graphene systems. The Al$_{2}$O$_{3}$ (0001) surface
is modeled by a slab containing six oxygen O$_{3}$ layers and twelve
or eleven aluminum layers (depending on the specific surface
studied). The second surface of the slab is passivated by pseudo H
atoms and a 15 \AA-vacuum region is included. Single-layer or
bilayer graphene with 2 $\times$ 2 periodicity is placed on 1
$\times$ 1 cell of Al$_{2}$O$_{3}$ with a lattice mismatch
$\sim$2\%\cite{strain-effect}. A $\Gamma$-centered 24 $\times$ 24
$\times$ 1 \textbf{k}-point sampling is used for the Brillouin-zone
integration. The energy cutoff is set to 400 eV and structural
optimization is carried out on all systems until the residual forces
are converged to 0.01 eV/\AA. The dipolar correction has been
included\cite{Neugebauer-1992}.

\begin{figure}[tbp]
\includegraphics[width=12.0cm]{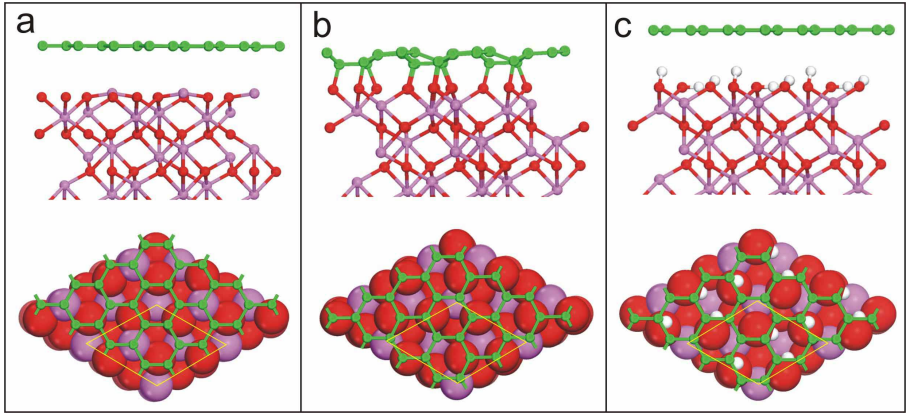}
\caption{Side and top views of optimized structures of graphene on
(a) Al-terminated, (b) O-terminated, and (c) fully hydroxylated
Al$_{2}$O$_{3}$ (0001) surface. The green, red, and pink balls
denote the C, O, and Al atoms, respectively. The yellow lines
represent the unit cell in calculations.}
\end{figure}

$\alpha$-Al$_{2}$O$_{3}$ has rhombohedral symmetry \emph{R}$\bar 3$c
($D_{3d}^6$, No. 167). The oxygen atoms form a nearly hcp structure,
and the metal atoms fill two-thirds of the octahedral sites between
the oxygen layers\cite{Henrich-1994}. The (0001) surface of
$\alpha$-Al$_{2}$O$_{3}$ is of major technology importance because
it is often used as a substrate in growth of semiconducting as well
as superconducting materials. Low-energy electron diffraction (LEED)
revealed that the (0001) crystal surface exhibits a 1 $\times$ 1
structure below $\sim$1250$^{\circ}$ in air or in
vacuum\cite{Henrich-1994, G. Renaud-1998, J. Toofan-1998}. Although
theoretical calculations demonstrated that Al-terminated (0001)
surface is the most stable one\cite{Felice-1999, Wang-2000}, both
Al- and O- terminated (0001) surfaces are observed in experiments
and considered in our present work\cite{Henrich-1994, G.
Renaud-1998, J. Toofan-1998}.

For the Al-terminated (0001) surface, our calculations show that the
topmost Al atoms move down $\sim$ 0.65 \AA~into the next oxygen
layer after the relaxation compared to the cleaved surface, as shown
in Fig. 1a. This surface reconstruction stabilizes the surface by
the large suppression of surface polarization\cite{Felice-1999,
Wang-2000}. In order to investigate the stable interface structures,
we have calculated all the high symmetrical arrangements between
graphene and Al-terminated surface, where a surface Al or O atom is
directly below either: a graphene C atom (T$_{Al}$ or T$_{O}$), the
hollow site of graphene C atoms (H$_{Al}$ or H$_{O}$), or the center
of C-C bridge site (B$_{Al}$ or B$_{O}$). In addition, the case of a
C atom directly above the center of surface Al-O bridge site is also
considered. Our DFT+D/PBE total energy calculations indicate that
the most stable configuration is the T$_{Al}$ (Fig. 1a) with an
adsorption energy of -76.4 meV/C-atom and the distance between
graphene layer and substrate surface is $\sim$ 2.74 \AA. The
adsorption energy of the T$_{Al}$ configuration is more negative
than other high symmetrical configurations by 2 $\sim$ 9 meV/C-atom.
It should be noticed that the vdW correction is very important in
this case: without the vdW correction, the calculated interface
distance of the T$_{Al}$ configuration would be drastically
increased to 3.31 \AA~ and the adsorption energy would be only -2.6
meV/C-atom; the energy difference between T$_{Al}$ and other
configurations is within 2 meV/C-atom in the absence of vdW
correction. LDA calculation could give a similar interlayer distance
between graphene and substrate surface, but the binding energy is
still largely underestimated by 34 meV/C-atom.

On the O-terminated surface, as each O atom has one dangling bond,
the O-terminated surface is chemically reactive and could strongly
interact with the graphene layer\cite{L. Han-2010}. For the lowest
energy configuration, all three surface O atoms in the unit cell
form chemical bonds with the graphene layer to suppress the surface
dangling bonds, as shown in Fig. 1b. The C-O bond length ranges from
1.45 to 1.50 \AA. These C-O binding severely distort the planar
graphene structure (C-C bond length varies from 1.36 to 1.49 \AA)
and the binding energy between graphene and O-terminated surface is
-349 meV/C-atom, which is much larger than that of Al-terminated
surface. In addition, fully hydroxylated Al$_{2}$O$_{3}$ surface is
also considered, as H atoms can be incorporated into the bulk
structure during growth\cite{Henrich-1994}. Previous calculations
also demonstrated that the stability of fully hydroxylated surface
is comparable to the Al-terminated one and much more stable than
O-terminated surface\cite{Felice-1999, Wang-2000}. Similar to
Al-terminated surface, graphene layer adsorbed on hydroxylated
surface also belongs to vdW interaction. The most stable
configuration is B$_{O}$ (one surface O directly below the center of
C-C bridge site), as shown in Fig. 1c. The interlayer distance
between graphene layer and hydroxylated surface ranges from 2.31
\AA~ to 3.18 \AA~due to the rippled structure at the interface and
the binding energy is -49.2 meV/C-atom, slightly smaller than the
binding strength of interlayers in graphite. The binding energy of
B$_{O}$ case is slightly lower than that of T$_{O}$ and H$_{O}$
within 1 meV/C-atom, indicating that these configurations are almost
degenerate in energy.

\begin{figure}[tbp]
\includegraphics[width=10.0cm]{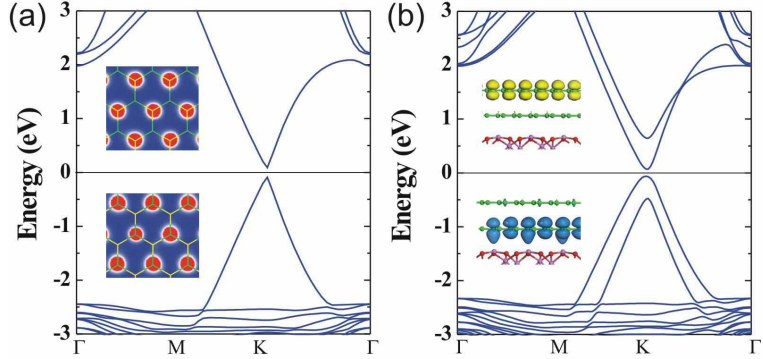}
\caption{HSE functional calculated electronic band structures of (a)
single-layer graphene and (b) bilayer graphene on Al-terminated
(0001) Al$_{2}$O$_{3}$ surface. The Fermi level is set to zero. The
partial charge densities of VBM and CBM for the two structures
are plotted as bottom and upper insets in (a) and (b),
respectively.}
\end{figure}

After knowing the stable interface structures, we turn to understand
the effects of Al$_{2}$O$_{3}$ substrate on the electronic property
of graphene. The HSE band structure of single-layer graphene
adsorbed on Al-terminated surface (T$_{Al}$ configuration) is shown
in Fig. 2a. Notably, the degeneracy of Dirac cone of graphene is
lifted and a band gap of 182 meV (69 meV in DFT+D/PBE calculations)
appears at the Fermi level, generating an effective mass of the
Dirac fermions $\sim$ 8 $\times$ 10$^{-3}$ m$_{e}$. This is because
the inversion symmetry in the graphene plane is broken on
Al$_{2}$O$_{3}$ surface so the A, B sublattices, which is equivalent
in free-standing graphene, is no longer equivalent. We find that the
electrostatic potential difference of graphene C atoms at the A and
B site differ by $\sim$ 165 meV, close to the value of the band gap.
The charge density distributions of conduction band minimum (CBM)
and valence band maximum (VBM) (inset of Fig. 2a) are found
separately located in the different A and B sublattices of graphene
layer, which is consistent with our explanation.

The sizes of band gaps of graphene in other high symmetrical
configurations are close to that of T$_{Al}$, indicating that a
similar ON/OFF ratio could be obtained even if graphene slides away
from its ground state T$_{Al}$ configuration. Remarkably, the band
gap of graphene on Al-terminated Al$_{2}$O$_{3}$ is comparable to
10-nm-width GNRs in experiments\cite{Wang-2008, M. Y. Han-2007, X.
Li-2008}. Meanwhile, the effective mass of carriers in
graphene/Al$_{2}$O$_{3}$ is an order lower than that of GNRs ($\sim$
0.07 m$_{e}$) from theoretical calculation\cite{M. Long-2009},
strongly indicating the much higher carrier mobility of
graphene/Al$_{2}$O$_{3}$ in practice. The band alignment between the
graphene and substrate surface is found to be type-I and there are
no interface states around Fermi level for the Al-terminated
interface (Fig. 2a). By distinguishing the respective graphene and
Al$_{2}$O$_{3}$ states, we can calculate the band offset for this
hybrid system. The valence band offset between graphene and
Al$_{2}$O$_{3}$ is about 2.35 eV from the hybrid functional
calculation (DFT+D/PBE level calculation seriously underestimates
the valence band offset by 1 eV), comparable to that of silicon and
Al$_{2}$O$_{3}$\cite{Xiang-2009}, which is high enough for injection
barrier.

The electronic properties of bilayer graphene on Al-terminated
surface, as shown in Fig. 2b. The equilibrium graphene interlayer
distance with AB stacking is 3.18 \AA. Similar to single-layer
graphene, a band gap of 138 meV is induced in bilayer graphene due
to the inversion symmetry breaking. The VBM and CBM are contributed
by A$_{1}$ and B$_{2}$ sublattices (A$_{1}$ and B$_{1}$ belong to
the bottom layer while A$_{2}$ and B$_{2}$ belong to the top layer)
in different graphene layers, respectively, as shown in Fig. 2b. The
valence band offset between bilayer graphene and Al$_{2}$O$_{3}$ is
2.26 eV, slightly smaller than that of single-layer graphene. It is
important to notice that the band gaps of single-layer or bilayer
graphene on Al-terminated surface are large enough to create
sufficient ON/OFF ratio in FET applications\cite{Qimin-2007, M. Y.
Han-2007, X. Li-2008, Wang-2008}. As we expected, the band gap of
graphene on Al$_{2}$O$_{3}$ is significantly larger than that on
other popular substrates such as SiO$_{2}$, HfO$_{2}$, and BN,
indicating that $\alpha$-Al$_{2}$O$_{3}$ is a better choice for gate
material\cite{note1-LDA}.

\begin{figure}[tbp]
\includegraphics[width=8.0cm]{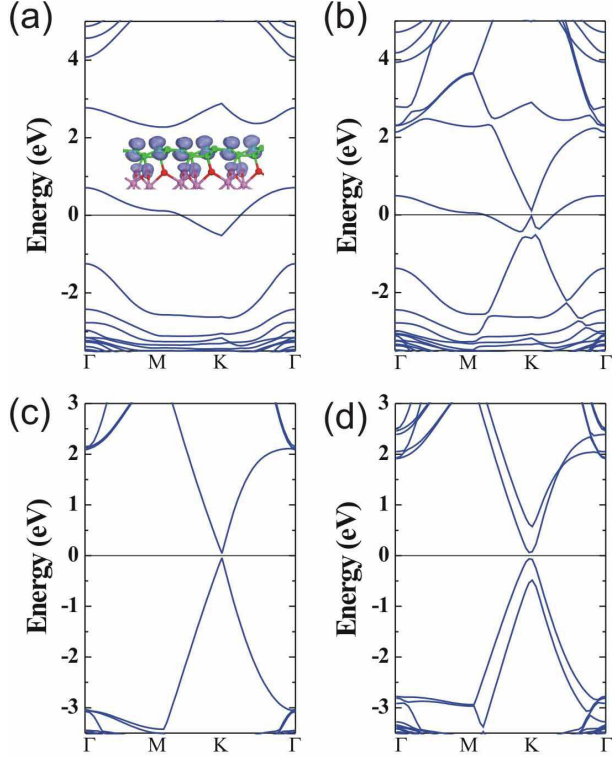}
\caption{HSE functional calculated electronic band structures of (a)
single-layer graphene and (b) bilayer graphene on O-terminated
(0001) Al$_{2}$O$_{3}$ surface; the partial charge densities of the
partially occupied band at the Fermi level in (a) are plotted as
inset. HSE electronic band structures of (c) single-layer graphene
and (d) bilayer graphene on fully hydroxylated (0001)
Al$_{2}$O$_{3}$ surface are also plotted. The Fermi level is set to zero.}
\end{figure}

The electronic properties of graphene on O-terminated surface are
completely different from that of Al-terminated one. Due to the
strong interaction between graphene and surface O atoms, the linear
band characteristic of graphene disappears, as shown in Fig. 3a. The
gap states in Fig. 3a originate from the hybridization between
graphene C orbitals and surface O orbitals, shown as the inset in
Fig. 3a. When a second graphene layer is placed on the surface (the
graphene interlayer distance varies from 3.09 \AA~to 3.75 \AA~due to
the rippled structure of the bottom layer), these gap states couple
with the second layer graphene states and perturb its linear band
distribution, as shown in Fig. 3b. A band gap of 140 meV appears for
the second graphene layer, but the interface states exist inside the
gap. Because these localized interface states can largely suppress
the high mobility of graphene, O-terminated surface must be avoided
to contact with graphene when gating graphene in experiments.

The electronic properties of graphene adsorbed on hydroxylated
surface show similar behaviors to the Al-terminated surface, as
shown in Figs. 3c and 3d for B$_{O}$ configuration. No interface
states around Fermi level are found and a band gap of 84 meV is
induced in graphene layer with the effective mass of $\sim$ 4
$\times$ 10$^{-3}$ m$_{e}$ for the Dirac fermion. The valence band
offset between graphene and hydroxylated surface is 3.39 eV, larger
than that of Al-terminated one. Moreover, the electronic properties
of T$_{O}$ and H$_{O}$ configurations are quite similar to that of
B$_{O}$. Comparing to the single layer case, the bilayer graphene on
the hydroxylated surface has a larger band gap of 126 meV and
smaller band offset of 3.24 eV, as shown in Fig. 3d. Clearly,
O-terminated surface should be hydroxylated in order to make clean
interface between graphene and Al$_{2}$O$_{3}$. Although the
amorphous structure of Al$_{2}$O$_{3}$ has been selected as a top
gate in some experiments\cite{Liao-Kim}, it is difficult for us to
directly compare our results with these experiments due to the
different phases.

Finally, we considered the electronic properties of
graphene/Al$_{2}$O$_{3}$ hybrid system under an external electric
field ($E_{ext}$) to simulate the gating effect in
experiments\cite{Neugebauer-1992}. Taking graphene on Al-terminated
surface as examples, Fig. 4 shows the band gap and valence band
offset of graphene/Al$_{2}$O$_{3}$ system as a function of
$E_{ext}$. For single-layer graphene on Al$_{2}$O$_{3}$, a negative
$E_{ext}$ increases the band gap while the trend is opposite for a
positive $E_{ext}$, as shown in Fig. 4a. This is because a positive
(negative) $E_{ext}$ increases (decreases) the distance between
graphene and surface, e.g., the distance increases (decreases) from
2.74 \AA~to 2.86 (2.69) \AA~when $E_{ext}$ = 0.6 (-0.6) V/\AA. The
increase of the distance between graphene and substrate reduces the
potential difference $\triangle_{AB}$, thus changing the band gap.
The variation of potential difference also influences the valence
band offset, as shown in Fig. 4. The valence band offset increases
(decreases) with a negative (positive) $E_{ext}$ and varies from
2.49 eV to 2.05 eV under -0.6 eV/\AA $\leq$ $E_{ext}$ $\leq$ 0.6
eV/\AA.

\begin{figure}[tbp]
\includegraphics[width=14.0cm]{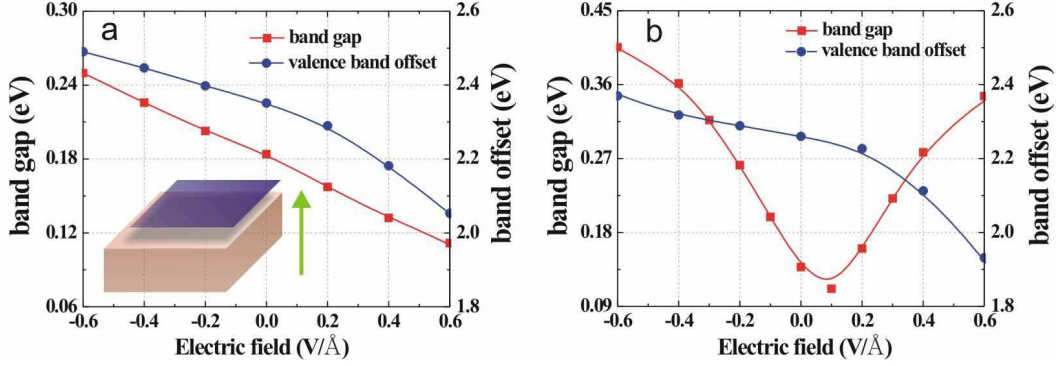}
\caption{HSE functional calculated band gap and valence band offset
of (a) single-layer graphene and (b) bilayer graphene on
Al-terminated Al$_{2}$O$_{3}$ surface as a function of external
electric field ($E_{ext}$). The principal scheme of the
computational model is shown as inset of (a). An $E_{ext}$ is
oriented normally to a surface and is assumed to be positive when it
is directed upward.}
\end{figure}

The situation is quite different for bilayer
graphene/Al$_{2}$O$_{3}$ (Fig. 4b) under an $E_{ext}$. Due to the
symmetry, the band gap of bilayer graphene is determined mainly by
the absolute difference of $|\triangle_{A_{1}B_{2}}|$.\cite{Y.
Zhang-2009} This value reaches minimum at $E_{ext} \sim 0.1$ V/\AA~
when the system has a minimum anticrossing band gap around 112 meV.
As $E_{ext}$ increases or decreases from this value,
$|\triangle_{A_{1}B_{2}}|$ increases, which makes the band gap
increases as well, as shown in Fig. 4b. For example, the band gap of
bilayer graphene increases to 361 meV at $E_{ext}$ = -0.4 V/\AA. The
valence band offset changes between 2.37 eV and 1.93 eV for -0.6
eV/\AA $\leq$ $E_{ext}$ $\leq$ 0.6 eV/\AA. Thus, even under high
$E_{ext}$, the valence band offset (injection barrier) of
graphene/Al$_{2}$O$_{3}$ is still large enough.

with physical insights and by first-principles calculations we
demonstrate that $\alpha$-Al$_{2}$O$_{3}$ is an ideal high-\emph{k}
gate material for graphene FETs. There are no interface states
between graphene and Al-terminated or hydroxylated Al$_{2}$O$_{3}$
surface and the valence band offset for this hybrid system is larger
than 2 eV. Remarkably, a band gap of $\sim$ 180 meV appears in
graphene layer deposited on Al-terminated surface, which is
significantly larger than on other popular used substrates.
Furthermore, the band gap of graphene/Al$_{2}$O$_{3}$ system could
be tuned by an external electric field for practice applications.

The authors thank the helpful discussion with Prof. Jaejun Yu (SNU).
The work at NREL was supported by the U.S. Department of Energy
under Contract No. DE-AC36-08GO28308.

\newpage

\end{document}